\documentclass[aps,10pt,twocolumn,showpacs,showkeys,pra,citeautoscript,amsmath,amssymb,floatfix,nofootinbib,superscriptaddress,longbibliography]{revtex4-2}
\usepackage[left=2cm,right=2cm,top=2cm,bottom=2cm]{geometry}

\usepackage[utf8]{inputenc}
\usepackage[english]{babel}
\usepackage[T1]{fontenc}
\usepackage{xcolor}

\usepackage{orcidlink}
\usepackage{cleveref} 
\usepackage{mathtools}
\usepackage{verbatim}
\usepackage{tikz-cd} 
\usepackage{amsmath}
\usepackage{amsfonts}
\usepackage{amssymb}
\usepackage{cancel}
\usepackage{braket}
\usepackage{array}   
\newcolumntype{L}{>{$}l<{$}} 
\usepackage{bbm}

\usepackage{bbold}
\newcommand{\id}[0]{\mathbb{1}}

\DeclareMathOperator{\tr}{tr}

\usepackage{soul}
\usepackage{amsthm}

\usepackage{enumerate}
\usepackage{braket}

\usepackage{framed}
\usepackage{graphicx}

\usepackage{natbib}

\usepackage{csquotes}

\usepackage{xcolor}
\usepackage[normalem]{ulem}
\usepackage{soul}

\newcommand{\blk}{\color{black}}

\usepackage{hyperref}

\begin{document}
\title{
How Quantum Contextuality disappears in the Classical Limit}
\author{Arthur C. R. Dutra*\orcidlink{0009-0000-2293-4929}}
\affiliation{Instituto de Matemática, Estatística e Computação Científica,
Universidade Estadual de Campinas, Campinas, SP, Brazil}
\author{Roberto D. Baldij\~ao\orcidlink{0000-0002-4248-2015}}
\affiliation{Perimeter Institute for Theoretical Physics, 31 Caroline Street North, Waterloo, Ontario N2L 2Y5, Canada}

\author{Marcelo Terra Cunha\orcidlink{0000-0002-8990-5329}}
\affiliation{Instituto de Matemática, Estatística e Computação Científica,
Universidade Estadual de Campinas, Campinas, SP, Brazil}

\begin{abstract}
The emergence of classicality is fundamentally driven by the interaction between a quantum system and its environment. Foundational open-system approaches, notably the Caldeira-Leggett model, successfully captured how these interactions lead to macroscopic effects like quantum dissipation and decoherence.
However, these approaches often leave the precise definitions of classicality and quantumness ambiguous.
In quantum information theory, this boundary is a heavily scrutinized question, and Kochen-Specker contextuality emerges as a hallmark of nonclassicality.
It is therefore natural to investigate whether decoherence can actually suppress this property.
Taking this path creates an apparent conundrum, once there exist two distinct manifestations of quantum contextuality: state-dependent and state-independent ones. 
While state-dependent contextuality naturally vanishes under state degradation, state-independent contextuality could persist for any quantum state, since it shows up even for the maximally mixed state!
In this paper, we resolve this apparent paradox by analyzing sequential measurement implementations of the paradigmatic  Klyachko, Can, Binicioğlu,
and Shumovsky (KCBS) and Peres-Mermin prepare-and-measure scenarios under the influence of depolarizing channels.
By introducing noise both prior to and in between measurements, and by analyzing the resulting sequential correlators in both the Schrödinger and Heisenberg pictures, we show how open-system dynamics suppress the correlations required to witness contextuality, leading to classicalization. 
\end{abstract}

\maketitle
\onecolumngrid

\twocolumngrid
{
\renewcommand{\thefootnote}%
    {\fnsymbol{footnote}}
  \footnotetext[1]{arthur.couto.oliveira@alumni.usp.br}
}

\section{Introduction}

The Caldeira-Leggett model is a cornerstone of the Classical Limit of Quantum Theory~\cite{CL-Model,CL-BM}.
By adapting the friction model from classical mechanics, Amir Caldeira and his PhD supervisor offered to the community a central tool to enforce Zeh's dictum ``there is no quantum isolated system''~\cite{giulini2013decoherence}. One way of introducing the decoherence paradigm is to say that
quantum systems classicalize through interaction with (somehow classical) environmental systems. 
However, when we express these ideas in this way, it seems that we have already agreed upon what it means for a system to be quantum or classical.
This sounds like an overly optimistic starting point, which we shall avoid.

Under the paradigm of seeking classicality as an emergent property from within quantum theory, we adopt the following criterion: a system may be regarded as classical whenever the statistics of all relevant measurements can be reproduced from a single joint probability distribution over their outcomes, thus providing a self-consistent classical description~\cite{Fine82}.
In classical statistical physics, such a global description always exists: different physical properties are assumed to possess well-defined values simultaneously, and their measurement statistics arise from a common underlying probabilistic model. Whenever this is not possible, genuinely nonclassical features are revealed. 

Let us start with an example to show when a system cannot be considered classical:
We can use the Clauser-Horne-Shimony-Holt (CHSH) Bell scenario, where two distant agents may each choose independently between two possible dichotomic measurements~\cite{CHSH,RevModPhys.38.447}. 
In this case, the demand for a classically consistent description is equivalent to asking for the obedience of every CHSH-Bell inequality~\cite{Fine82,TerraBundles}.
Hence, a quantum system capable of generating a violation of a Bell inequality must not be considered  classical (e.g. it can be used to generate quantum advantage in a communication task like quantum key distribution~\cite{PhysRevLett.67.661}).
Violation of Bell inequalities can be seen as a gold standard of a non-classicality marker.

Bell Nonlocality can be thought of as just the tip of an iceberg called Kochen-Specker (KS) contextuality~\cite{kochen1990problem,budroni2021quantumReview}. At its core, KS contextuality arises from 
a central novelty of quantum theory:  the existence of incompatible measurements (classically, any two measurements that make sense individually also make sense jointly).
This naturally leads to the notion of a \emph{context}, namely, a set of mutually compatible measurements. A KS scenario is then specified by a set of measurements  together with their  compatibility relations~\cite{BarbaraBook}. The assumption of noncontextuality requires that the value associated with a measurement be independent of the context in which it appears. Equivalently, the statistics observed in all \emph{contexts} must be reproducible from a single joint probability distribution over the outcomes of all measurements in the scenario~\cite{fine-theorem}. This matches the classical view that different physical properties possess well-defined values simultaneously. Whenever this is not possible, the scenario exhibits contextuality, which can also be witnessed through the violation of noncontextuality inequalities.

A bipartite Bell scenario is an example of a KS scenario where compatibility relations are ensured by spatial separation:
if $\left\{ A_i\right\}$ and $\left\{B_j\right\}$ are, respectively, the sets of possible measurements available for Amir and Bell, the set of contexts will be given by the pairs $\left\{ (A_i,B_j)\right\}$. In the more general KS-contextuality scenarios, however, neither two parties nor spacelike separation are required to reveal nonclassicality~\cite{budroni2021quantumReview}. A single observer, Amir alone in his lab, may already obtain statistics that do not admit a noncontextual (classical) description. 

As will be better described in Sec.~\ref{Sec:KS}, Bell scenarios generate examples of state-dependent contextuality.
As the name suggests, in those cases, the non-classicality marker, given by the violation of some noncontextuality inequality, depends on the quantum state of the system (e.g. separable states can not generate any violation of  a Bell Inequality).
Classical limit in those cases can not only be achieved, but also interpreted in a very comfortable way: decoherence process drives the initial state, which was capable of showing some violation, into a mixed state from which every correlation admit classical explanation~\cite{Baldijao_2020_StateDependentMultiple,baldi_thesis}.   

As convincing as this interpretation may seem, it opens another intriguing question. There are also examples of \emph{state-independent contextuality}~\cite{PhysRevLett.101.210401}, namely, proofs of nonclassicality involving sets of measurements for which even the maximally mixed state generates correlations that do not admit a classical description (see Sec.~\ref{Sub:SIC}). Do they also exhibit a classical limit?
A negative answer to this question would mean the prevalence of  nonclassical properties over decoherence, which could challenge the idea that classicality emerges from  decoherence processes.
The positive answer brings us another question: how to understand it if not by the degradation of the quantum state?

In this paper, we show that state-independent contextuality also undergoes a classical limit, marked by the vanishing of violations of KS noncontextuality inequalities.
Before describing these results, we will start by a generous introduction to Kochen-Specker contextuality, in Sec.~\ref{Sec:KS}, emphasizing its two manifestations and how it is the central aspect of nonclassicality. 
{In Sec.~\ref{sec:Methods} we review the methods we use, in particular introducing quantum channels and their connection to the usual approaches to open systems.}
Sec.~\ref{Sec:Emergence} brings the technical part of the work,  reviewing the previous results for state-dependent cases before addressing the state-independent one.
We close the article with some discussions in Sec.~\ref{Sec:Disc}. 

\section{Kochen-Specker contextuality}\label{Sec:KS}

As argued in the introduction, we consider a system to admit a classical description when its properties can be modeled in a self-consistent way.
Let us justify this statement.

One of the central novelties of quantum theory is the existence of incompatible  measurements, related to observables that cannot be jointly measured. 
One could think that incompatibility alone would imply nonclassicality.
However, the mere empirical limitation of not being able to measure two incompatible, yet repeatable, observables simultaneously does not necessarily mean a classical description is impossible.
Instead, a classical mindset might attempt to reconcile this by postulating that, although experimentally inaccessible, an overarching global probabilistic description of all possible measurements exists, and the experiments we can perform simply reveal a restricted part of it. For example, Spekkens Toy Model~\cite{Spekkens_2007} is a model which has incompatibility due to an epistemic restriction, but does admit a classical explanation.  

To make the distinction between nonclassical and classical scenarios with incompatibility precise, consider a quantum system and a set of repeatable measurements $\mathcal{M} = \{A_1, A_2, A_3, \dots\}$. 
A subset of \emph{compatible} measurements define a context, here denoted $C$.
We will denote by $a_k$ the outcome obtained by measuring $A_k$ and by $P(a_i, a_j, \dots | A_i, A_j, \dots)$ the outcome probability distribution of a joint measurement of the context $C=\{A_i,A_j,\dots\}$. 
The classical (usually tacit, but here explicit) assumption of  \textit{ noncontextuality} posits the existence of a single global joint probability distribution, $P(a_1, a_2, a_3, \dots|A_1, A_2, A_3, \dots)$, over all measurements in $\mathcal{M}$, regardless of their mutual (in)compatibility \cite{budroni2021quantumReview}. 

Crucially, if this classical description holds, $P(a_1, a_2, a_3, \dots|A_1, A_2, A_3, \dots)$ must act as the overarching distribution for all observable phenomena. The empirical probability of obtaining specific outcomes within an experimentally accessible context $C= \{A_i, A_j, \dots\}$ must simply be the marginalization over the unmeasured (usually incompatible) observables' outcomes:
\begin{multline}
 P(a_i, a_j, \dots | C) = \blk \\
    \sum_{a_k \,:\, A_k \notin C} P(a_1, a_2, \dots | A_1, A_2, \dots).
\end{multline}
If such a global probability distribution exists, the system's correlations admit a classical explanation. Indeed, the probability distribution associated with any experimentally accessible context is obtained by marginalizing one and the same underlying distribution, regardless of the context considered. This is precisely the sense in which the model is \emph{noncontextual} and this is what we mean by a classical self-consistent account~\cite{AbramskyBrandemburger_2011}.

The Kochen-Specker (KS) theorem~\cite{kochen1990problem}, however, demonstrates that this classical reconciliation may fail for quantum theory.
It proves that experiments using quantum systems do not, in general, admit a global joint probability distribution. 
Therefore, the statistics of compatible joint measurements cannot be derived from a single context-independent underlying distribution. This inescapable context-dependence when attempting to complete quantum theory is one of the distinguished non-classical features of the quantum world.

While the failure of the marginal problem elegantly defines KS contextuality, it poses a practical challenge: one cannot experimentally assess the \textit{non-existence} of an inaccessible global distribution directly. 
This is how noncontextuality inequalities save the day.
Formulated as a generalization of Bell inequalities~\cite{PhysicsPhysiqueFizika.1.195,CHSH}, they establish a rigorous mathematical equivalence between the existence of a global joint probability distribution and the satisfaction of a finite set of linear inequalities by the empirically accessible outcome probability distributions~\cite{ Fine82}\footnote{At least for contextuality scenarios made of finite numbers of measurements and outcomes are considered.}.
In practice, it means that an empirical behavior is contextual if and only if it violates at least one noncontextuality inequality. \blk
These inequalities are typically expressed as bounds on a linear combination of the expectation values of the joint measurements within the allowed contexts. 
For a given set of contexts, $\mathcal{C}=\{C_k\}_k$,  a generic noncontextuality inequality may be written as
\begin{equation}\label{eq:genNCineq}
    \sum_{k} \gamma_k \langle C_k \rangle \stackrel{NC}{\leq} \beta_C,
\end{equation}
where $\langle C_k \rangle$ represents the expectation value of the joint measurement associated with context $C_k=\{A_i, A_j, \dots\}$ and \enquote{NC} stands for \enquote{noncontextual}, as this inequality emerges from the classical noncontextual description.
This expectation value is given by the sum of the product of the individual outcomes weighted by their joint probability:
\begin{equation}
    \langle C_k \rangle = \sum_{a_i, a_j, \dots} (a_i a_j \cdots) P(a_i, a_j, \dots | A_i, A_j, \dots).
\end{equation}
Furthermore, the $\gamma_k$ are real coefficients defining the specific inequality, and $\beta_C$ is the classical bound---the maximum value achievable by any noncontextual (classical) model. By experimentally estimating these expectation values, one can directly witness contextuality, i.e., the impossibility of a classical, self-consistent, explanation.

So far, we have emphasized the properties of the observables and their compatibility structures. 
However, to fully characterize KS contextuality, we must also consider the state in which the quantum system was prepared. 
The interplay between the measurements and the preparation brings forth an important distinction in contextuality phenomena: whether the contextual behavior is tied to specific states or if it manifests universally for all states. We will explore this distinction in the following examples.

\subsection{State-dependent contextuality}\label{Sub:SDC}
State-dependent contextuality is conceptually the closest relative to Bell scenarios. 
Indeed, Bell nonlocality gives the mostly known cases of state-dependent contextuality and possibly the reader has already recognized CHSH inequality as an example of expression \eqref{eq:genNCineq}. 
In a typical Bell test, whether or not we observe correlations that exceed classical limits strongly depends on the prepared state. 
Changing the state to something noisy will generally wash away the non-local signature. 
Similarly, in this form of contextuality, the violation of noncontextual bounds is not a universal property of the measurements themselves, but rather a joint feature of good choices of  state and measurements.
\label{exmp: KCBS}

The most famous example of indivisible state-dependent contextuality is the KCBS scenario, originally introduced in Ref.~\cite{PhysRevLett.101.020403}, which consists of five dichotomic measurements arranged in a cyclic compatibility structure:
\begin{align}
\begin{array}{@{}c@{\;}l@{}}
\begin{array}{c}
K\\[-2.5pt]
C\\[-2.5pt]
B\\[-2.5pt]
S
\end{array}
&
\begin{cases}
\mathcal{M}=\{A_0,A_1,A_2,A_3,A_4\};\\
\mathcal{O}=\{-1,1\};\\
\mathcal{C}=
\left\{
\begin{aligned}
&\{A_0,A_1\},\{A_1,A_2\}, \{A_2,A_3\},\\
&\{A_3,A_4\}, \{A_4,A_0\}
\end{aligned}
\right\}.
\end{cases}
\end{array}
\end{align}
This scenario exhibits the following noncontextuality inequality \cite{PhysRevLett.101.020403}\footnote{We keep it in the original format, but it can be recognized as an NC-inequality \eqref{eq:genNCineq} with all $\gamma_k=-1$ and $\beta_C = 3$.
Indeed, this scenario is completely characterized by equivalent inequalities \cite{PhysRevA.88.022118}.}:
\begin{equation}
    \langle A_0A_1\rangle+\langle A_1A_{2}\rangle+\langle A_2A_{3}\rangle+\langle A_3A_{4}\rangle+\langle A_4A_{0}\rangle\stackrel{NC}{\geq} -3.
    \label{KCBS inequality}
\end{equation}

A quantum realization of this scenario is possible in a three dimensional quantum system, with the observables
\begin{equation}
    A_i=2\ket{v_i}\bra{v_i}-\id,
\end{equation}
where $\id$ is the identity matrix and 
\begin{equation}
    \ket{v_i}=\left(\frac{1}{\sqrt[4]{5}}, \sqrt{1-\frac{1}{\sqrt{5}}}\cos{\left(\frac{i\pi4}{5}\right)} , \sqrt{1-\frac{1}{\sqrt{5}}}\sin{\left(\frac{i\pi4}{5}\right)} \right)^T.
\end{equation}

This realization exhibits state-dependent contextuality because only \textbf{some} quantum states violate the inequality. 
For instance, by preparing the state $\ket{\psi}\bra{\psi}$, with
\begin{equation}
    \ket{\psi}=    \begin{pmatrix} 1 \\ 0 \\ 0 \end{pmatrix},
\end{equation}
we can get the maximum quantum violation given by
\begin{equation}
    \sum_{i=0}^{n-1}\langle A_i,A_{i+1}\rangle_{\ket{\psi}\bra{\psi}}\stackrel{Q}{=}5-4\sqrt{5}<-3,
\end{equation}
where \enquote{Q} stands for \enquote{quantum}, as this equality emerges from the quantum description. 

Experimental detections of KS contextuality with a quantum realization of the KCBS scenario were made  using single photons 
\cite{lapkiewicz2011experimental}, ions \cite{um2013experimental,malinowski2018probing} and superconductors~\cite{Jerger2016}.
Much like in Bell nonlocality, however, if the prepared state is sufficiently mixed, its behavior may no longer produce a violation. An extreme example is the maximally mixed state, $\rho = \id/3$, which only achieves a value of $\sum_i \langle A_i A_{i+1}\rangle = -5/3$, falling well within the classical noncontextual region bounded by $-3$.

\subsection{State-independent contextuality}\label{Sub:SIC}
\label{sec:SIC}
Perhaps surprisingly, KS contextuality also admits state-\textit{independent} forms. In these scenarios, the impossibility of a noncontextual description is rooted in the measurement structure itself, namely the (in)compatibility relations among the jointly measurable observables, giving rise to noncontextuality inequalities that are universally violated, even by the maximally mixed state. 
This phenomenon has no analog in standard Bell scenarios, where locality constraints can always be satisfied by sufficiently noisy preparations, since entanglement is a necessary condition and a fragile resource.

A famous example of state-independent contextuality is the Peres-Mermin (PM) square, presented in references \cite{PERES1990107,PhysRevLett.65.3373,APeres_1991}. As a contextuality scenario, it is defined by a set of nine dichotomic measurements and six contexts
\begin{align}
\begin{array}{@{}c@{\;}l@{}}
\begin{array}{c}
P\\[-2.5pt]
M
\end{array}
&
\begin{cases}
\mathcal{M}=\{A_{11},A_{12},A_{13},A_{21},A_{22},A_{23},A_{31},A_{32},A_{33}\};\\
\mathcal{O}=\{-1,1\};\\
\mathcal{C}=\bigl\{\{A_{i1},A_{i2},A_{i3}\},\{A_{1i},A_{2i},A_{3i}\}\bigr\}_{i\in\{1,2,3\}}.
\end{cases}
\end{array}
\end{align}
which can be placed in a square grid
\begin{align}
\begin{split}
\label{eq:PM_square}
    &A_{11} \hspace{0.3cm} A_{12} \hspace{0.3cm} A_{13} \\
    &A_{21} \hspace{0.3cm} A_{22} \hspace{0.3cm} A_{23} \\
    &A_{31} \hspace{0.3cm} A_{32} \hspace{0.3cm} A_{33}, \\
\end{split}
\end{align}
such that each row and column of the grid forms one of the contexts.

This scenario has the following KS-noncontextuality inequality \cite{PhysRevLett.101.210401}\footnote{The inequality presented in Ref.~\cite{PhysRevLett.101.210401} is slightly different from Ineq. \eqref{square inequality}. One can be obtained from the other through relabeling \(A_{11}\rightarrow-A_{11}\) and \(A_{12}\rightarrow-A_{12}\).}:
\begin{align}
\begin{split}
    &\langle A_{11}A_{12}A_{13} \rangle +\langle A_{21}A_{22}A_{23}\rangle +\langle A_{31}A_{32}A_{33} \rangle\\ 
-&\langle A_{11}A_{21}A_{31}\rangle-\langle A_{12}A_{22}A_{32}\rangle -\langle A_{13}A_{23}A_{33}\rangle\stackrel{NC}{\leq} 4,
\end{split}
\label{square inequality}
\end{align}
and can be realized in a two-qubit quantum system with the Pauli measurements:
\begin{align}
\begin{split}
A_{11}&=\sigma_y\otimes \sigma_z \hspace{0.3 cm} A_{12}=\sigma_z\otimes \sigma_y \hspace{0.3 cm}
A_{13}=\sigma_x\otimes \sigma_x \\
A_{21}&=\sigma_z\otimes \sigma_x \hspace{0.3 cm} A_{22}=\sigma_x\otimes \sigma_z \hspace{0.3 cm}
A_{23}=\sigma_y\otimes \sigma_y \\
A_{31}&=\sigma_x\otimes \sigma_y \hspace{0.3 cm} A_{32}=\sigma_y\otimes \sigma_x \hspace{0.3 cm}
A_{33}=\sigma_z\otimes \sigma_z.
\label{squareMeasurements}
\end{split}
\end{align}

Through quantum theory we can calculate the expected value of each term of inequality \eqref{square inequality} by multiplying the observables of each context. 
The ones from each row of \eqref{squareMeasurements} multiply to $\id$ while those from the columns multiply to $-\id$.
This means that, for any quantum state $\rho$, quantum theory predicts that the left-hand side of inequality \eqref{square inequality} must give
\begin{align}
6\langle \id \rangle=6,
\label{quantum equality}
\end{align}
irrespectively of which state is used to take the mean value,
which shows that this realization exhibits state-independent contextuality.

Researchers have experimentally detected KS contextuality in quantum realizations of the PM scenario using ions \cite{kirchmair2009state}, nuclear magnetic resonance \cite{PhysRevLett.104.160501}, and photons \cite{PhysRevLett.103.160405}. 
 Interestingly, even though significant violations of the classical bound are obtained in all those realizations, the left hand side of inequality~\eqref{square inequality} is usually also inconsistent with the expected quantum value $6$. 
 Such inconsistency is usually attributed to experimental imperfections, but this paper brings  a refined view of this apparent inconsistency. \blk

\subsection{The compatibility vs individual existence loophole}
\label{sec:CompatibilityLoophole}
While the mathematical frameworks of both state-dependent and state-independent contextuality clearly demonstrate a fundamental departure from classicality, there is one final physical subtlety we must discuss. 

To experimentally demonstrate a violation of these inequalities, one must implement the observables in different measurement contexts. In any contextuality scenario, the assumption of noncontextuality requires that, when $A_i$ appears in the contexts as $\{A_i,A_j\}$ and $\{A_i,A_k\}$, it corresponds to the same physical measurement in both cases. 
If, however, each context is implemented through a single, monolithic joint-measurement device, one may question whether the physical identity of $A_i$ is truly preserved across different contexts. Since changing the context amounts to changing the entire apparatus, a critic may argue that the observable being measured is not operationally the same object in all cases. 
Peres-Mermin square inequality \eqref{eq:PM_square} is a good example: no one would consider six measurements of the identity operator as a contextuality test: the individual measurements $A_{ij}$ are demanded.
This is the so-called individual existence loophole~\cite{individual_existence}. 

A natural way to address this objection is to adopt a sequential measurement protocol on a single quantum system. In this approach, a context $C=\{A_i,A_j\}$ is not measured by a single black-box device, but by sending the system through distinct measuring devices in succession. The system first interacts with the apparatus associated with $A_i$, producing an outcome $a_i$, and is then directed to the apparatus associated with $A_j$, producing an outcome $a_j$. Because the same physical device is used to implement $A_i$ independently of whether the subsequent measurement is $A_j$ or $A_k$, the operational identity of the observable is much more transparent. 

At the same time, this operational choice introduces another subtlety. In a sequential implementation, one must additionally assume that the first measurement does not disturb the statistics of the second. Otherwise, an apparent violation could be attributed not to contextuality itself, but to residual measurement disturbance, giving rise to the compatibility loophole~\cite{compatibility_loophole}. Nevertheless, when one must choose between these two experimental imperfections, the sequential arrangement is the implementation most frequently implemented in experimental settings. In this paper, we will analyze contextuality scenarios assuming sequential measurements.

The central point for the present work is that the sequential solution also introduces an unavoidable operational constraint: time must pass between successive measurements. 
As the system travels from the first apparatus to the second, it cannot remain perfectly isolated, and will in general interact with its surrounding environment.  
Thus, the  same arrangement that makes measurement identity clear also opens the door to decoherence. It is precisely this interplay between sequential measurements, environmental disturbance, and the progressive emergence of classical behavior that we will formalize in the following sections.

\section{Methods: decoherence through quantum channels}
\label{sec:Methods}

To address the emergence of the classical limit we must consider the system of interest  not as an isolated token, but as an open quantum system interacting with an environment. 
Historically, there are two canonical frameworks to describe this general non-unitary dynamics: the description of time-evolution generators and the characterization via quantum channels.

The first approach focuses on the continuous time evolution of the density matrix $\rho(t)$. 
In the Markovian limit, the dynamics are governed by the Gorini-Kossakowski-Sudarshan-Lindblad master equation~\cite{env_deoherence}:
\begin{equation}
    \frac{d\rho}{dt} = -i[H, \rho] + \sum_k \gamma_k \left( L_k \rho L_k^\dagger - \frac{1}{2}\{L_k^\dagger L_k, \rho\} \right),
    \label{eq:Lindblad}
\end{equation}
where $H$ is the Hamiltonian and $L_k$ are the Lindblad operators describing the coupling to the environment. 

While the Lindblad equation is essential for continuous-time dynamics, the second approach, via quantum channels, is often more advantageous for Quantum Information tasks where operations are discrete.  In this framework, any physical process that transforms a state $\rho$ into $\rho'$ is described by a Quantum Channel, a linear map $\mathcal{E}$ that is Completely Positive and Trace-Preserving (CPTP)~\cite{nielsen_chuang_2010}.
The action of such a channel can be written in terms of Kraus operators $\{K_i\}$:
\begin{equation}
    \mathcal{E}(\rho) = \sum_i K_i \rho K_i^\dagger,
    \label{eq:Kraus_map}
\end{equation}
subject to the completeness relation $\sum_i K_i^\dagger K_i = \id$, which ensures the preservation of the trace.

A crucial concept for our analysis of contextuality, particularly for state-independent scenario, is the notion of the \textit{dual channel}. While the Schrödinger picture map $\mathcal{E}$ describes how states evolve (decohere), the Heisenberg picture map $\mathcal{E}^\dagger$ describes how observables are transformed by interaction. The duality is defined via the trace inner product 
\begin{equation}
\label{eq:dualMaps}
    \tr(A \mathcal{E}(\rho)) = \tr(\mathcal{E}^\dagger(A) \rho), \forall A,\rho.
\end{equation}
It follows that the dual map is given by $\mathcal{E}^\dagger(A) = \sum_i K_i^\dagger A K_i$. 
It is important to note that $\mathcal{E}$  trace-preservingness corresponds to $\mathcal{E}^\dagger$ being unital (i.e., $\mathcal{E}^\dagger(\id) = \id$). 
By duality, the Heisenberg channel $\mathcal{E}^\dagger$ is trace-preserving if, and only if, the Schrödinger channel is unital ($\sum_i K_i K_i^\dagger = \id$).

To make the discussion concrete, we now introduce a paradigmatic family of channels associated with a  decoherence model that will be used throughout the remainder of the paper: the depolarizing channels. 
These channels capture the loss of coherence that typically occurs due to interaction with an isotropic environment, and they admit a simple description in arbitrary finite dimensions.

For a $d$-dimensional quantum system, a depolarizing channel characterized by a parameter $0\leq p\leq 1$ is defined as
\begin{align}
\label{eq:DepolarizingChannel}
\mathcal{D}_p(\rho) = p \rho + (1-p)\frac{\id}{d}.
\end{align}
Eq.~\eqref{eq:DepolarizingChannel} can be interpreted as: the channel does nothing to the system with probability $p$, and re-prepares the system in the maximally mixed state with probability $1-p$. For $p=1$, the map reduces to the identity channel, corresponding to noiseless and trivial evolution. For $p=0$, one obtains the totally depolarizing channel,
\begin{align}
\mathcal{D}_0(\rho) = \frac{\id}{d},
\end{align}
which maps every input state to the maximally mixed state. The parameter $p$ therefore quantifies how much of the original state survives the interaction with the environment. Increasing the strength or duration of the system–environment coupling typically drives $p$ toward zero, corresponding to increasing loss of coherence.
It is important to note that for such a channel (and all unital channels, indeed), there is no role for the energetic hierarchy of ground and excited states.

A particularly important feature of this channel is its simple action in the Heisenberg picture. Applying the defining duality relation in Eq.~\eqref{eq:dualMaps}
one finds that the dual map acts as
\begin{align}
\label{eq:DualDep}
\mathcal{D}^\dagger_p(A)
=
pA + (1-p)\frac{\tr(A)}{d}\,\id.
\end{align}
In particular, for any traceless observable $A$ (like Pauli matrices),
\begin{align}
\mathcal{D}^\dagger_p(A) = pA.
\end{align}
Thus, in the Heisenberg picture, depolarizing noise uniformly contracts all traceless observables by the same factor $p$. This simple ``shrinking'' property will be central in our later analysis of state-independent contextuality under decoherence.

For concreteness, let us briefly specialize  to the qubit case. For $d=2$, the depolarizing channel becomes
\begin{align}
\label{eq:DepQubit}
\mathcal{D}_p(\rho) = p\rho + (1-p)\frac{\id}{2}.
\end{align}
The qubit example allows us to visualize the action of $\mathcal{D}_p$: it shrinks the Bloch Ball, such that the depolarized states live in a ball with radius $p$ after its action.
A convenient Kraus decomposition is given in terms of the four ``Pauli'' operators $\id$, $\sigma_x,\sigma_y,\sigma_z$, with equal strength to the three sigmas and a different one for the identity:
\begin{subequations}    
\begin{align}
K_0 &= \sqrt{\kappa}\,\id, \\
K_i &= \sqrt{\iota}\,\sigma_i,
\qquad i=x,y,z,
\end{align}
\end{subequations}
which yields
\begin{align}
\label{eq:KraussDepQubit}
\mathcal{D}_p(\rho)
=
\kappa \rho
+
\iota
\sum_{i=x,y,z}
\sigma_i \rho \sigma_i.
\end{align}
In order to recover Eq.~\eqref{eq:DepQubit}, one can recognize that 
\begin{equation}
    \sum_{\mu = 0,x,y,z} \sigma_{\mu} \rho \sigma_{\mu} = 2\id = 4 \frac{\id}{2},
\end{equation}
and conclude that $\kappa = \displaystyle \frac{1+3p}{4}$ and $\iota = \displaystyle \frac{1-p}4$ do the job.
This shows that a depolarizing map $\mathcal{D}_p$ can equivalently be understood as: with probability $p$, nothing happens, while with probability $\frac{1-p}{4}$, the state gets one of the four possible `Pauli kicks', even if the identity `Pauli kick' can not be distinguished from doing nothing (and their probabilities add).
Let us now see how the qubit depolarizing channel arises naturally from a Markovian open-system dynamics. Consider, for instance, the Lindblad equation
\begin{align}
\frac{d\rho}{dt}
=
\gamma
\sum_{i=x,y,z}
\left(
\sigma_i \rho \sigma_i - \rho
\right),
\end{align}
where $\gamma$ quantifies the strength of the system--environment interaction. 
This is a master equation with Lindblad operators $L_i=\sqrt{\gamma}\,\sigma_i$ and $H=0$.
Physically, this generator describes isotropic, unital, Markovian noise acting equally along all Bloch-sphere directions, so that no ``spatial'' direction is privileged and the Bloch vector contracts uniformly toward the maximally mixed state. 
Such a situation arises, for instance, for a spin-$1/2$ particle coupled isotropically to three independent bosonic environments through an interaction Hamiltonian of the form $\displaystyle H_{\mathrm{int}}=\sum_{i=x,y,z}\sigma_i \otimes B_i$. Under the usual Born--Markov and secular approximations, and assuming identical bath correlation functions in all directions, the reduced dynamics of the spin takes precisely the form above.
A simpler realization is the dynamics of photonic polarization crossing an isotropic depolarizing medium.

Writing the state in Bloch form, $\rho = (\id + \mathbf{r}\cdot\boldsymbol{\sigma})/2$, one finds that the Bloch vector obeys $\dot{\mathbf{r}} = -4\gamma\,\mathbf{r}$, so that
\begin{align}
\rho(t)
=
e^{-4\gamma t}\rho(0)
+
\left(1-e^{-4\gamma t}\right)\frac{\id}{2}.
\end{align}
The reduced dynamics therefore implement a depolarizing channel with a time-dependent parameter
\begin{align}
p(t) = e^{-4\gamma t}.
\end{align}
In other words, a depolarizing channel $\mathcal{D}_p$ is implemented for each value of $(\gamma,t)$. Increasing either the coupling strength $\gamma$ or the interaction time $t$ drives the system toward the maximally mixed state, corresponding to stronger decoherence.

While we have illustrated the connection to Lindblad equations and Kraus operators explicitly for qubits, the defining form $\mathcal{D}_p(\rho)=p\rho+(1-p)\id/d$ and the corresponding dual action extend straightforwardly to arbitrary finite dimensions. 
In what follows, we will use this channel as a model for decoherence\footnote{Naturally $\mathcal{D}_p$ is not the only channel that captures decoherence. For instance, the dephasing map would also represent  a decoherence process in which a pointer basis is perfectly selected by the environment.}: in the Schrödinger picture it drives states toward the maximally mixed state, while in the Heisenberg picture it uniformly suppresses traceless observables. These two complementary viewpoints will allow us to track how contextuality witnesses degrade under open-system dynamics.
\blk

\section{Emergence of Noncontextuality as Classical Limit}\label{Sec:Emergence}
This is the central section of the paper.
In Subsec.~\ref{Sub:NcycleClassicalLimit}we will apply the  techniques above to consider the impact of decoherence in the KCBS scenario, showing how the classical limit is attained for these cases of state-dependent contextuality. 
We close the section with the richer situation of the Peres-Mermin square, which demands a different understanding of how decoherence leads to classicalisation in this case.

\subsection{State-dependent contextuality: KCBS}\label{Sub:NcycleClassicalLimit}

We now address the question: if decoherence is regarded as a classicalization process, does it make contextuality disappear, so that the result of the process becomes consistent with noncontextuality?

To analyze this question, consider a test of state-dependent contextuality implemented after the action of a noisy channel. As a concrete example, we examine the Klyachko–Can–Binicioğlu–Shumovsky (KCBS) inequality for a qutrit system (see Sec.~\ref{exmp: KCBS}). The KCBS inequality reads
\begin{align}
    \sum_{i=0}^{4} \langle A_i A_{i+1} \rangle \ge -3,
\end{align}
where indices are taken modulo 5, and the observables $A_i$ are dichotomic and satisfy the compatibility relations defining the 5-cycle scenario.

Suppose the system is prepared in a state $\rho$ that provides a maximal quantum violation of the KCBS inequality, and assume that, before the measurements are performed, the system interacts with an environment. If $\mathcal{E}$ denotes the channel describing this interaction, we wish to determine whether the evolved state $\mathcal{E}(\rho)$ still violates the inequality. In other words, even if $\rho$ violates the KCBS inequality, it is not guaranteed that $\mathcal{E}(\rho)$ will do so.

Let us begin with the extremal case of a completely depolarizing channel. In dimension $d=3$, this channel maps any state to the maximally mixed state $\id/3$. For the KCBS observables one finds
\begin{align}
    \langle A_i A_{i+1} \rangle_{\rho=\frac{\id}{3}}
    =
    -\frac{1}{3},
    \qquad \forall i,
\end{align}
so that
\begin{align}
    \sum_{i=0}^{4}
    \langle A_i A_{i+1} \rangle_{\rho=\frac{\id}{3}}
    =
    -\frac{5}{3}.
\end{align}
Since $-\frac{5}{3} > -3$, the maximally mixed state does not violate the KCBS inequality. Thus, perfect depolarization necessarily erases any state-dependent contextuality. Once complete decoherence has occurred, the system cannot exhibit the nonclassical correlations present prior to the interaction.

We now turn to the more realistic case of partial depolarization. Consider the depolarizing channel introduced in Sec.~III A,
\begin{align}
    \mathcal{D}_p(\rho)
    =
    p \rho + (1-p)\frac{\id}{3},
\end{align}
with $0 < p \le 1$. 
One way of realizing such channel is to leave the input state unchanged with probability $p$ and to replace it by the maximally mixed state with probability $1-p$. 
The smaller the value of $p$, the stronger the loss of coherence.

Using the linearity of expectation values with respect to the state, we obtain
\begin{align}
    \sum_{i=0}^{4}
    \langle A_i A_{i+1} \rangle_{\mathcal{D}_p(\rho)}
    &=
    p \sum_{i=0}^{4}
    \langle A_i A_{i+1} \rangle_{\rho}
    +
    (1-p)
    \sum_{i=0}^{4}
    \langle A_i A_{i+1} \rangle_{\frac{\id}{3}} \nonumber \\
    &=
    p S_\rho
    -
    (1-p)\frac{5}{3},
\end{align}
where we have defined
\begin{align}
    S_\rho :=
    \sum_{i=0}^{4}
    \langle A_i A_{i+1} \rangle_{\rho}.
\end{align}

The state $\mathcal{D}_p(\rho)$ violates the KCBS inequality if and only if
\begin{align}
    p S_\rho - (1-p)\frac{5}{3} < -3.
\end{align}
Solving this inequality for $p$, and assuming $S_{\rho} < -3$, we obtain
\begin{align}
    p > 
    \frac{-3+\frac{5}{3}}
    {S_\rho+\frac{5}{3}}
    =
    \frac{-\frac{4}{3}}
    {S_\rho+\frac{5}{3}},
\end{align}
which defines
\begin{align}
\label{eq:pcritical}
    p_{\rm crit}(\rho)
    :=
    \frac{-\frac{4}{3}}
    {S_\rho+\frac{5}{3}}.
\end{align}
We therefore conclude that the state $\mathcal{D}_p(\rho)$ violates the KCBS inequality only if $p > p_{\rm crit}$. Equivalently, no violation occurs whenever $p \le p_{\rm crit}$. This condition holds for any state $\rho$ that initially violates the inequality\footnote{Note that, for every violating state, one has $S_\rho < -3$, which implies that $S_\rho + \frac{5}{3} < 0$. Hence the denominator in Eq.~\eqref{eq:pcritical} is negative for violating states, and since the numerator $-\frac{4}{3}$ is also negative, it follows that $ p_{\rm crit}(\rho) >0$.} and subsequently undergoes depolarization with parameter $p$.

To determine the strongest possible threshold, we consider the state that achieves the maximal quantum violation of the KCBS inequality, for which $S_\rho = 5 - 4\sqrt{5}$. Substituting this value into the expression for $p_{\rm crit}$ yields
\begin{align}
p_{\rm crit}^{\rm max}=\max_{\rho}p_{\rm crit}(\rho) = \frac{5 + 3\sqrt{5}}{20} {\approx 0.59}.
\end{align}
Thus, no state — including the one achieving maximal quantum violation — violates the KCBS inequality after passing through a partially depolarizing channel with
$p \le \frac{5 + 3\sqrt{5}}{20}.$
Notice that the same result for $p_{\rm crit}^{\rm max}$ can equivalently be obtained in the Heisenberg picture, using the dual map $\mathcal{E}_{p}^{\dagger}$ together with Eqs.~(20) and (21), and the fact that $\tr[A_{i}A_{i+1}] = -1$ for the KCBS observables.

We conclude that state-dependent contextuality is fragile under decoherence: even if the initial state exhibits maximal quantum violation, a sufficiently strong interaction with an environment depolarizing the state suppresses the correlations below the noncontextual bound. In this sense, decoherence drives the system towards classical behavior.

This conclusion is consistent with other scenarios in which contextuality is degraded through interaction with additional degrees of freedom. In particular, in Ref.~\cite{Baldijao_2020_StateDependentMultiple} we showed that when a single system is sequentially probed by many independent observers, the accumulation of measurement-induced disturbance similarly suppresses contextual correlations. In both cases, interaction with external systems — whether modeled as an environment or as additional observers — leads to the disappearance of contextuality.

\subsection{State-independent contextuality: the Peres-Mermin square}\label{Sub:PMClassicalLimit}

Following the logic of the previous subsection, one might be tempted to conclude that decoherence should not destroy state-independent contextuality. In the state-dependent case, we can explain the classical limit emerging due to the noisy evolution degrading the initial state towards one that no longer violated the relevant inequality. But this reasoning cannot be applied here, since state-independent contextuality survives even for the maximally mixed state. If decoherence were understood only as a process that erases special features of the preparation, there would be no obvious reason to expect the violation to disappear. This could suggest, at least at first glance, that state-independent contextuality could be immune to the usual classicalizing effect of noise, despite the fact that nothing like this seems to persist in our classical experience of the world.

One avenue for solving this tension is to note that, so far, we have modeled decoherence as acting on the system after preparation and before the contextuality test, that is, before any of the measurements. However, as discussed in Sec.~\ref{sec:CompatibilityLoophole}, contextuality tests designed to avoid the individual existence loophole are implemented through sequential measurements, which necessarily involve time passing between the measurements of a context. During this interval, the system cannot remain perfectly isolated. This opens a second route through which decoherence may degrade contextual correlations: by acting in-between compatible measurements.

One interesting way to begin this analysis is to rewrite the depolarizing channel in terms of the Pauli twirl
\begin{align}
    \begin{split}
        \mathcal{D}_p(\rho)
        &= p\rho + (\id-p)\,\frac{I}{4},\\
        &= p\rho + \frac{1-p}{16}\sum_{i,j\in\{x,y,z,I\}}\,\boldsymbol{\sigma}_{ij} \,\rho\, \boldsymbol{\sigma}_{ij},
    \end{split}
\end{align}
where $\boldsymbol{\sigma}_{ij} \coloneqq\sigma_i \otimes \sigma_j$ denotes the two-qubit Pauli operator, with $\sigma_x,\, \sigma_y,\, \sigma_z$ being the usual Pauli matrices and $\sigma_I$ representing the $2\times 2$ identity matrix. This is a two-qubit generalization of Eq.~\eqref{eq:KraussDepQubit}.

This means that applying noise to the state is equivalent to, with a probability $1-p$, having an additional hidden observer implementing a random Pauli measurement before the desired one\footnote{You can read a full description of this interpretation in Ref.~\cite{public_systems}}. This formulation already hints at the problem: since the Pauli measurements are not all compatible, we would expect that introducing a random incompatible measurement in the scenario would somehow interfere with the violation. In fact, since any given Pauli product commuted with 8 of the Paulis products and anti-commutes with the other 8, the second term cancels when applying the adjoint to a Pauli pair, so we get a shrinking effect
\begin{align}
    \begin{split}
    \mathcal{D}_p^\dagger(\boldsymbol{\sigma}_{ij})
&= p\,\boldsymbol{\sigma}_{ij}
    \end{split}
\end{align}
for any Pauli pair that is not the identity, $\boldsymbol{\sigma}_{II}$. This should also hint at what is happening, since these depolarized Pauli measurements can become jointly measurable \cite{PhysRevD.33.2253,Leo}, and collective joint measurability cannot exhibit contextuality.

To precisely determine how this shrinking factor propagates through the full evaluation of the Peres-Mermin inequality
\begin{align}
\begin{split}
    &\langle A_{11}A_{12}A_{13} \rangle +\langle A_{21}A_{22}A_{23}\rangle +\langle A_{31}A_{32}A_{33} \rangle\\ 
-&\langle A_{11}A_{21}A_{31}\rangle-\langle A_{12}A_{22}A_{32}\rangle -\langle A_{13}A_{23}A_{33}\rangle\stackrel{NC}{\leq} 4,
\end{split}
\label{square inequality 2}
\end{align}
we must model how the sequential measurement process affect the nine measurements from the square:
\begin{align}
\begin{split}
\label{eq:PM_square-bis}
    &A_{11} \hspace{0.3cm} A_{12} \hspace{0.3cm} A_{13} \\
    &A_{21} \hspace{0.3cm} A_{22} \hspace{0.3cm} A_{23} \\
    &A_{31} \hspace{0.3cm} A_{32} \hspace{0.3cm} A_{33}, \\
\end{split}
\end{align}

To compute the expectation values associated with the row and column contexts, we model each projective measurement through its corresponding Lüders instrument. For a dichotomic observable $A_{ij}$ with outcomes $\pm1$, we define
\begin{equation}
\label{eq:score_map_projective}
\mathcal{S}_{ij}(\rho)
:=
\Pi^{ij}_{+}\rho\Pi^{ij}_{+}
-
\Pi^{ij}_{-}\rho\Pi^{ij}_{-},
\end{equation}
where
\begin{equation}
\Pi^{ij}_{\pm}=\frac{\id\pm A_{ij}}{2},
\qquad
A_{ij}=\Pi^{ij}_{+}-\Pi^{ij}_{-},
\qquad
A_{ij}^2=\id.
\end{equation}
The expectation value of a single measurement is then
\begin{equation}
\langle A_{ij}\rangle=\tr\!\left[\mathcal{S}_{ij}(\rho)\right]
=\tr(A_{ij}\rho).
\end{equation}
Likewise, the correlators entering the row and column contexts are obtained by composing the corresponding Lüders maps in sequence.
For instance, for two sequentially measured compatible observables \(A_{ij}\) and \(A_{kl}\),
\begin{equation}
\langle A_{ij}A_{kl}\rangle
=
\tr\!\left[\mathcal{S}_{kl}(\,\mathcal{S}_{ij}(\rho))\right],
\end{equation}
and analogously for longer sequences  involving compatible measurements.

Let us now consider a sequential measurement of one of the PM contexts, which we assume, without loss of generality, to be one of the rows. We denote by \(\langle A_{i1}A_{i2}A_{i3}\rangle_{\mathcal{D}_p^{\times 3}}\)
the expectation value obtained when the context
\((A_{i1},A_{i2},A_{i3})\) is implemented sequentially, with the noise channel \(\mathcal{D}_p\) acting on the system immediately before each measurement. With the definition of the channel \(\mathcal{S}_{ij}\), this expectation value can be written in the Schrödinger picture as
\begin{widetext}
\begin{equation}
\label{eq:schrodinger_triple_projective}
\langle A_{i1} A_{i2} A_{i3}\rangle_{\mathcal{D}_p^{\times 3}}
=
\tr\!\left[
\mathcal{S}_{i3}\!\Big(
\mathcal{D}_p\!\big(
\mathcal{S}_{i2}\!\big(
\mathcal{D}_p\!\big(
\mathcal{S}_{i1}\!\big(
\mathcal{D}_p(\rho)
\big)\big)\big)\big)\Big)
\right].
\end{equation}
The Schrödinger picture calculation is cumbersome for this scenario, which can obscure intuition about the process. By invoking the duality of quantum channels ($\tr[A\mathcal{E}(\rho)] = \tr[\mathcal{E}^\dagger(A)\rho]$), we can move to the Heisenberg picture . Here, the dynamics are effectively pulled back onto the identity operator at the end of the sequence, allowing us to analyze the evolution of the observables directly:
\begin{align}
\label{eq:heisenberg_triple_nested}
&\langle A_{i1} A_{i2} A_{i3}\rangle_{\mathcal{D}_p^{\times 3}}=
\tr\!\left[
\rho\;
\mathcal{D}_p^\dagger\!\left(
\mathcal{S}_{i1}^\dagger\!\left(
\mathcal{D}_p^\dagger\!\left(
\mathcal{S}_{i2}^\dagger\!\left(
\mathcal{D}_p^\dagger\!\left(
\mathcal{S}_{i3}^\dagger(\id)
\right)\right)\right)\right)\right)\right].
\end{align}
To solve this, we rely on two main properties. First, the noise channel shrinks any Pauli observable $\boldsymbol{\sigma}_{ij}$ by $p$ but leaves the identity invariant:
\begin{equation}
\label{eq:Ep_adj_props}
\mathcal{D}_p^\dagger(\id)=\id,
\qquad
\mathcal{D}_p^\dagger(A_{ij})=p\,A_{ij}.
\end{equation}
Second, since the measurements within a context commute, the Heisenberg map for a measurement simply multiplies the current operator by the measured observable:
\begin{align}
\label{eq:commuting_pushthrough}
\mathcal{S}_{i3}^\dagger(\id)=A_{i3},\quad
\mathcal{S}_{i2}^\dagger(A_{i3})=A_{i2}A_{i3},\quad
\mathcal{S}_{i1}^\dagger(A_{i2}A_{i3})=A_{i1}A_{i2}A_{i3}.
\end{align}
We can now evaluate the nested sequence in Eq. \eqref{eq:heisenberg_triple_nested} step-by-step from the inside out.
\begin{align}
\begin{split}
\mathcal{D}_p^\dagger\!\left(
\mathcal{S}_{i1}^\dagger\!\left(
\mathcal{D}_p^\dagger\!\left(
\mathcal{S}_{i2}^\dagger\!\left(
\mathcal{D}_p^\dagger(A_{i3})
\right)\right)\right)\right)&=
\mathcal{D}_p^\dagger\!\left(
\mathcal{S}_{i1}^\dagger\!\left(
\mathcal{D}_p^\dagger\!\left(
\mathcal{S}_{i2}^\dagger(pA_{i3})
\right)\right)\right)\\
&=
\mathcal{D}_p^\dagger\!\left(
\mathcal{S}_{i1}^\dagger\!\left(
p\,\mathcal{D}_p^\dagger(A_{i2}A_{i3)}
\right)\right) \\
&=
\mathcal{D}_p^\dagger\!\left(
p^2\,A_{i1}A_{i2}A_{i3}
\right)\\
&=
p^2\,A_{i1}A_{i2}A_{i3},
\label{eq:Kp_p2},
\end{split}
\end{align}
where, from the second to the third line, we used that the product $A_{i2}A_{i3}$ always results in a Pauli observable $\mathbf{\sigma}_{ij}$ and the last step is a result of the three observables multiplying to the identity.
\end{widetext}
Thus, the white noise shrinks the expected value of the joint measurement by a factor of $p^2$
\begin{subequations}
\label{eq:NoisyCorrelatorsPM}
\begin{equation}
    \langle A_{i1} A_{i2} A_{i3}\rangle_{\mathcal{D}_p^{\times 3}} = p^2\langle A_{i1} A_{i2} A_{i3}\rangle,
\end{equation}
and likewise for the column contexts 
\begin{equation}
    \langle A_{1j} A_{2j} A_{3j}\rangle_{\mathcal{D}_p^{\times 3}} = p^2\langle A_{1j} A_{2j} A_{3j}\rangle.
\end{equation}
\end{subequations}

Since the classical bound for the inequality in Expression~\eqref{square inequality 2} is $4$, whenever $p^2 \leq 2/3$ the noise suppresses the expected values below the classical bound, and no violations are witnessed for the state-independent inequality. This shrinking mechanism may also help to explain a persistent feature of Peres--Mermin experiments~\cite{PhysRevLett.103.160405,kirchmair2009state,Qu2021}: although modern implementations achieve impressive values around $5.7$ or $5.8$, these results are still not compatible with the ideal quantum value\footnote{In these experiments if we attributed all the error to such a depolarizing channel, it would correspond to a value of $p \approx 0.98$.} of $6$.

What we get from Eqs.~\eqref{eq:NoisyCorrelatorsPM} is the following: notwithstanding  the ideal Peres--Mermin construction being state-independent, its nonclassical signature need not remain observable if the sequential implementation is subjected to noise  in between measurements. 
When this effect is properly taken into account, we understand how the experiment may fail to witness nonclassicality even when the underlying ideal scenario is contextual, for sufficient noise. 

If one adopts the perspective suggested by the dual-map calculation, the effective measurements realized in the noisy sequential experiment are no longer the ideal sharp ones, but noisy versions of them. In that case, the traditional KS notion is no longer the most natural one for analyzing the observed statistics, since it is tied to ideal sharp measurements. This points instead toward Spekkens' generalized framework, which is better suited to noisy and unsharp measurement procedures and allows one to assess nonclassicality in a noise-robust way~\cite{Krishna_2017}. From this viewpoint, part of the noise can be incorporated into the description of the scenario itself, rather than treated only as a deviation from an ideal setting. Still, the main conclusion remains unchanged: sufficiently strong noise drives the experiment into a regime in which no nonclassicality can be witnessed.

\section{Discussion}\label{Sec:Disc}

In this paper, we explore a classical limit for KS contextuality. Adopting the viewpoint that a system admits a classical description whenever the observed statistics can be explained by a single joint probability distribution, we call systems for which this intuitive property cannot be achieved \emph{contextual}. From this perspective, any genuine classical limit must be accompanied by the disappearance of contextuality,
{a process that can fairly be called \emph{decontextualization}.}

To analyze this question, we modeled decoherence through depolarizing channels.
For state-dependent contextuality, like KCBS scenario, it is consistent to understand the process in the following way: the action of the channel drives the prepared state toward the maximally mixed state and progressively suppresses the violation of noncontextuality inequalities, eventually restoring compatibility with a classical description~\cite{GeometryESD}. 
In this case, the usual intuition that decoherence degrades the relevant nonclassical properties of the state is sufficient to explain the emergence of the classical limit.

The state-independent case is subtler. Since the corresponding violations persist even for the maximally mixed state, degradation present only at the preparation stage cannot explain how the classical limit emerges. Our analysis shows that the key point is that, in realistic sequential implementations of compatible measurements, noise may act not only before the contextuality test but also in between the ideal measurements that define each context. Using dual maps to obtain the values of the relevant correlators, we show that depolarization suppresses the observable signature of state-independent contextuality, to the point that the noisy sequential implementation may cease to witness any violation even when the underlying ideal scenario is contextual. Furthermore, our approach highlights how a fundamental hypothesis underpinning the assumption of KS noncontextuality, the requirement for sharp projective measurements, is difficult to strictly satisfy in these noisy scenarios. Consequently, it serves as a compelling example of the benefits of adopting the Spekkens approach to contextuality, particularly its noise-robust formulations~\cite{Krishna_2017}, when studying open quantum systems.

Our results therefore show that decoherence can account for a classical limit for both state-dependent and state-independent contextuality. It also points to 
a broader conceptual point: the Schr\"odinger-picture intuition, centered on the evolution of the quantum state, captures well the fate of state-dependent contextuality, but does not intuitively account for the state-independent case. There, the key insight comes not the state alone, but how the dynamics affect the correlations operationally accessible in a sequential scenario.
Although our treatment still follows a state-centered perspective, with the role of the dynamics made explicit only indirectly through the use of quantum channels and their dual maps, it point towards 
a viewpoint in which decoherence is understood as a process shaping the experimental scenario as a whole.

Decoherence is a subtle and important physical process.
Viewing it exclusively through a state-centric perspective does not pay the necessary tribute to its importance.
This hundred years of quantum theory and forty-five of Caldeira-Leggett model seem to be a great opportunity for reflection on whether the prevalence of the Schrödinger picture can divert us from the essential call from Heisenberg that the theory should focus its concerns on the observable entities\cite{heisenberg1925quantum}.

\section*{Acknowledgements}
It is natural to dedicate this paper to Amir Caldeira, who has strongly contributed to the growth of a quantum information community in Brazil.
We would also like to thank Rafael Rabelo and  Matthew Pusey for fruitful conversations.

\section*{Statements and Declarations}

\subsection*{Funding}
This study was financed in part by the Coordenação de Aperfeiçoamento de Pessoal de Nível Superior (CAPES)- Finance Code 001- by CNPq under grant 311314/2023-6, by the São Paulo Research Foundation (FAPESP) under the process numbers 2024/23590-2 and 2024/16657-3.
This work is part of the National Institute of Science and Technology of Quantum Information (INCT-IQ). 
RDB acknowledges support from Perimeter Institute. Research at Perimeter Institute is supported
in part by the Government of Canada through the
Department of Innovation, Science and Economic
Development and by the Province of Ontario
through the Ministry of Colleges and Universities. 

\subsection*{Competing Interests}
The authors have no competing interests to declare that are relevant to the content of this article.

\subsection*{Ethics Declaration}
Not applicable.

\bibliographystyle{unsrtnat}
\bibliography{Bib}

\appendix

\end{document}